\documentclass[aps,twocolumn,showpacs,preprintnumbers,floatfix]{revtex4}

\usepackage{mathrsfs}
\usepackage{amsmath}
\usepackage{graphicx}
\usepackage{subfigure}
\usepackage{epstopdf}

\begin{document}

\title{Baryons in QCD$_{\rm AS}$ at Large $N_c$: A Roundabout Approach}

\author{Thomas D. Cohen and Daniel L. Shafer}
\email{cohen@physics.umd.edu}
\email{dshafer@umd.edu}
\affiliation{Maryland Center for Fundamental Physics, Department of
Physics, University of Maryland, College Park, MD 20742-4111}

\author{Richard F. Lebed}
\email{Richard.Lebed@asu.edu}
\affiliation{Department of Physics, Arizona State University, Tempe,
AZ 85287-1504}

\date{December 2009}

\begin{abstract}
QCD$_{\rm AS}$, a variant of large $N_c$ QCD in which quarks transform
under the color two-index antisymmetric representation, reduces to
standard QCD at $N_c \! = \! 3$ and provides an alternative to the
usual large $N_c$ extrapolation that uses fundamental representation
quarks.  Previous strong plausibility arguments assert that the
QCD$_{\rm AS}$ baryon mass scales as $N_c^2$; however, the complicated
combinatoric problem associated with quarks carrying two color indices
impeded a complete demonstration.  We develop a diagrammatic technique
to solve this problem.  The key ingredient is the introduction of an
effective multi-gluon vertex: a ``traffic circle'' or ``roundabout''
diagram.  We show that arbitrarily complicated diagrams can be reduced
to simple ones with the same leading $N_c$ scaling using this device,
and that the leading contribution to baryon mass does, in fact, scale
as $N_c^2$.
\end{abstract}

\preprint{DOE-40762-475}
\preprint{INT-PUB-09-061}

%
% 11.15.Pg Expansions for large numbers of components (e.g.,
% 1/Nc expansions)
% 14.20.-c Baryons (including antiparticles)

\maketitle

\section{Introduction}

In 1974 't~Hooft proposed~\cite{'tHooft:1973jz} a generalization of
QCD from three to an arbitrary number $N_c$ of colors.  Many aspects
of QCD simplify in the large $N_c$ limit: Inasmuch as the $N_c \!
\rightarrow \! \infty$ and $N_c \! = \! 3$ worlds are qualitatively
similar, one can learn about QCD by starting with the large $N_c$
limit and expanding physical quantities systematically in powers of
$1/N_c$.  This approach has provided a powerful tool to study QCD and
other strongly coupled gauge theories.  It is implemented by
generalizing the gauge group of QCD with coupling $g$ from SU(3) to
SU($N_c$), and taking $N_c \! \rightarrow \! \infty$ while keeping
$g^2 N_c$ and the number of flavors $N_f$ fixed.  In this limit, each
quark loop is suppressed by a factor of $N_c^{-1}$, while each
nonplanar contribution to a diagram leads to suppression by a factor
of $N_c^{-2}$, and thus the number of diagrams one must consider is
radically reduced.  In 1+1 dimensions this diagrammatic simplification
is sufficient to allow direct calculation of the spectrum at leading
order in $1/N_c$~\cite{'tHooft:1974hx}, but unfortunately is not
sufficient in higher dimensions.  Nevertheless, by studying the $N_c$
scaling behavior of correlation functions, one can deduce the scaling
of various quantities with $N_c$.

The study of large $N_c$ baryons suffers the complication that the
physical operators acting upon them are not independent of $N_c$.
However, as observed by Witten~\cite{Witten:1979kh}, one can employ a
combinatoric analysis to deduce the scaling of observables associated
with baryons.  Moreover, a spin-flavor symmetry for baryons emerges at
large $N_c$, allowing one to relate various observables with relative
errors of $O(1/N_c)$, or in some cases
$O(1/N_c^2)$~\cite{emergent,DM,Dashen:1993jt,Carone:1993dz}.  These
relations tend to describe the real world quite well; for early
examples, see
Refs.~\cite{Dashen:1993jt,Dashen:1994qi,Jenkins:1994md,Luty:1994ub,
Jenkins:1995td,Dai:1995zg}.  Thus, large $N_c$ QCD and the $1/N_c$
expansion provide both qualitative insight and semiquantitative
predictions.

However, it has long been recognized that the extrapolation from
$N_c=3$ to large $N_c$ is not unique~\cite{Corrigan:1979xf}.  While
assuming that quarks at any $N_c$ transform in the fundamental (F)
representation of SU($N_c$) seems natural, it is not mandatory.
Corrigan and Ramond~\cite{Corrigan:1979xf} (CR) noted that using
quarks transforming in the two-index antisymmetric (AS) representation
of SU($N_c$) provides an equally valid extrapolation to large $N_c$.
At $N_c \! = \! 3$ the AS representation is just the $\overline{\bf
3}$, and thus simply swapping all AS quark and antiquark labels
reduces the theory to standard QCD\@.  The original motivation for
this construction was a scheme in which baryons have three quarks at
any $N_c$: one AS quark and two F quarks.

The study of large $N_c$ QCD containing quarks in the AS
representation has been revived in the past few years and has received
considerable attention.  The motivation is a profound new insight by
Armoni, Shifman, and Veneziano~\cite{Armoni:2003gp}, that an
``orientifold equivalence'' exists at large $N_c$ between QCD with
quarks in the AS representation and QCD with quarks in the adjoint.
One immediate consequence follows in the case of one massless quark
flavor: The equivalent theory is supersymmetric, and all of the
considerable formal power of SUSY can be brought to bear on the
problem.  A key issue is whether one can use this connection to learn
information---even approximate information---about QCD with more than
one flavor, as occurs in the physical world.

One strategy to deal with the real-world case of multiple flavors is
to focus on the chiral limit and allow one quark flavor to transform
in the AS representation while the other flavors transform as F, thus
maintaining supersymmetry at large $N_c$.  In point of fact, this
approach is just the CR scheme and gives 3-quark baryons at any $N_c$.
Unfortunately, it also suffers the major defect of breaking flavor
symmetry at any $N_c \! \neq \! 3$, which can then only be restored by
summing the all-orders expansion in $1/N_c$.  Since flavor is a
critical part of QCD dynamics, truncating the expansion at any low
order of $1/N_c$ is highly problematic.  This issue is discussed in
some detail in Ref.~\cite{Cherman:2009fh}.

Given this difficulty, one may consider a different implementation in
which {\em all\/} flavors of quark transform in the AS representation.
We refer to this theory as QCD$_{\rm AS}$ and distinguish it from the
usual theory QCD$_{\rm F}$ in which the quarks transform in the F
representation.  QCD$_{\rm AS}$ has the disadvantage of losing the
equivalence with a supersymmetric theory, but it correctly retains the
flavor symmetries, including chiral symmetry, that are essential to
reproduce correct low-energy QCD dynamics.  QCD$_{\rm AS}$ and
QCD$_{\rm F}$ coincide at $N_c \! = \! 3$, but extrapolate to large
$N_c$ in different ways; while the two large $N_c$ extrapolations have
much in common, they also differ in important respects.  While
nonplanar diagrams are suppressed in both the QCD$_{\rm F}$ and
QCD$_{\rm AS}$ limits, quark loops are not suppressed in the QCD$_{\rm
AS}$ limit because AS quarks, like gluons, carry two color indices.
This feature alters the nature of large $N_c$ scaling in the theory.

{\it A priori}, it is not obvious whether the $1/N_c$ expansion based
on QCD$_{\rm F}$ or the $1/N_c$ expansion based on QCD$_{\rm AS}$
gives a superior description of the $N_c \! = \! 3$ world.  Presumably
the answer depends upon which observable is considered.  A recent
comparison of baryon mass splittings~\cite{Cherman:2009fh} from the
two expansions suggests that {\em both\/} do a good job.

However, this comparison depends upon understanding the nature of
baryons in the two limits.  The case of baryons in QCD$_{\rm F}$ is
well known: The baryon mass was shown by Witten to scale as
$N_c^1$~\cite{Witten:1979kh}.  Witten's reasoning was somewhat
heuristic; it was based on a study of QCD with heavy
(non-relativistic) quarks and an argument that QCD with light quarks
should behave analogously.  A more formal version of this argument
valid for light quarks was developed in Ref.~\cite{Luty:1993fu}.  Both
of these treatments are based upon a diagrammatic analysis and thus
are not strictly rigorous; the analyses depend upon the assumption
that the $N_c$ scaling can be deduced from the contributions of
diagrams.  That is, they assume that fundamentally nonperturbative
effects do not alter the leading $N_c$ scaling.  However, up to this
assumption, the analysis is reliable.  The crux of the analysis is the
demonstration that the contribution to the mass due to the interaction
of $n$ quarks scales with $N_c$ in the same way as the contribution
from diagrams containing $n$ quarks, all of which are connected by
gluons, summed over all possible combinations of $n$ quarks (out of
$N_c$ total) that can contribute.

The second critical part of the analysis for QCD$_{\rm F}$ is the
demonstration that the contributions from $n$ connected quarks all
scale as $N_c^1$ or less; this demonstration is quite straightforward.
Consider the simplest case: one-gluon exchange between a pair of
quarks.  The contribution is proportional to $g^2 \sim N_c^{-1}$ times
an $O(N_c^2)$ combinatoric factor that specifies the number of
distinct quark pairs to which the gluons couple, giving a total of
$O(N_c^1)$.  It is easy to see that this is scaling is unaltered if
the two quarks are connected by a more complicated set of (planar)
gluon exchanges.  Moreover, it is also easy to see that the counting
is $O(N_c^1)$ for 3-quark interactions; in effect, adding a third
quark costs a factor of $N_c^{-1}$ due to the additional factor of
$g^2$ needed to couple it to the other quarks but brings in an
additional combinatoric factor of $N_c^1$.  This analysis can be
extended to the interactions of an arbitrary number of quarks, and for
all cases yields a leading scaling of $N_c^1$.
 
The situation for QCD$_{\rm AS}$ is more complicated.
Bolognesi~\cite{Bolognesi:2006ws} has shown that the operator creating
the lightest color-singlet state with nonzero baryon number in
QCD$_{\rm AS}$ with $N_c$ odd contains a product of $N_c(N_c-1)/2 \!
\sim \!  N_c^2$ quark operators, one carrying each color combination
$c_1 c_2$, where $c_1 \! \neq \! c_2$.  The associated states are
naturally identified with baryons.  This observation suggests that
baryon masses scale as $N_c^2$ instead of $N_c^1$, as in the case of
QCD$_{\rm F}$.  Such a conclusion seems particularly plausible since
Skyrme-type models in QCD$_{\rm AS}$, in which baryons are topological
solitons of QCD$_{\rm AS}$ meson fields, produce baryon masses scaling
as $N_c^2$~\cite{Cherman:2006iy}.  It is important to show that this
argument is more than plausible, and is in fact correct.  The natural
approach is simply to extend the analysis used in
Refs.~\cite{Witten:1979kh,Luty:1993fu} for QCD$_{\rm F}$ to the case
of QCD$_{\rm AS}$.  As noted above, this analysis contains two key
elements: The first is a demonstration that the contribution to the
mass due to the interaction of $n$ quarks scales with $N_c$ in the
same way as the contribution from diagrams containing $n$ quarks, all
of which are connected by gluons, summed over all possible
combinations of $n$ quarks (out of $N_c^2$ total) that can contribute;
this demonstration is easily extended to the case of QCD$_{\rm AS}$.
The second is an analysis of the scaling of contributions from $n$
connected quarks.  For the case of QCD$_{\rm AS}$ one expects the
scaling $\sim \! N_c^2$.  This second part is where the complexity
lies.

Indeed, as discussed in Ref.~\cite{Cherman:2006iy}, an apparent
paradox arises in computing the scaling of the contribution from $n$
connected quarks.  Consider, for example, the case of one-gluon
exchange between two quarks and use reasoning analogous to that of
Ref.~\cite{Witten:1979kh}.  Again, this contribution has two coupling
constants that contribute a scaling factor of $1/N_c$.  Again, one
computes a combinatoric factor: Assuming that the gluon can couple to
any of the $N_c^2$ quarks, one apparently obtains a factor of $N_c^4$,
yielding an overall contribution of $N_c^3$.  This result contradicts
the expectation that the mass scales as $N_c^2$.  Moreover, similar
reasoning leads one to conclude that the contribution of 3-quark
clusters scales as $N_c^4$, and more generally that $n$-quark clusters
scale as $N_c^{n+1}$.

The paradox is resolved by a more careful treatment of the
combinatorics.  The wave function of a QCD$_{\rm F}$ baryon consists
of a set of terms in which each color appears associated with
precisely one quark (with the full wave function fully antisymmetrized
in order to form a color singlet).  A gluon exchange between any two
quarks switches their colors, leading to a final state again
containing each color once, and again corresponding to a color
singlet.  Thus, the exchange of a gluon between {\em any\/} two quarks
maintains the color-singlet structure, and one obtains a combinatoric
factor given by the total number of pairs of quarks in the state.  In
contrast, consider what happens in QCD$_{\rm AS}$; in this case a
baryon again contains each color combination once, by which is meant a
pair of distinct fundamental color indices ({\it e.g.}, red-blue).  A
typical gluon exchange between two quarks exchanges {\em one\/} of the
two colors, yielding two quarks with different color combinations than
the initial two.  For example, consider two quarks $q_{a b}$ and $q_{c
d}$ where $a,b,c,d$ are the individual color indices, and assume that
all four of these colors are distinct (We use this specific example
later, in Sec.~\ref{sec:QCD_AS}).  Suppose a gluon exchange swaps two
of these colors (say $b$ and $c$); following this exchange the quarks
become $q_{a c}$ and $q_{b d}$, which are different from the original
quarks.  However, by construction these quarks inhabit a baryon in
which the color-singlet nature of the state requires that each color
combination occurs once and only once.  The exchange described above
is inconsistent with this constraint: After the exchange, one has two
copies of $q_{a c}$ and $q_{b d}$ and no copies of $q_{a b}$ or $q_{c
d}$.  Thus, in a color-singlet baryon the exchange described above
cannot occur.  In order for a one-gluon exchange diagram to appear
within a baryon state, it must occur between quarks that share a
color.  Thus, for example, the quarks $q_{a b}$ and $q_{c a}$ can
exchange a gluon that swaps $b$ and $c$, yielding $q_{a c}$ and $q_{b
a}$, consistent with the color-singlet constraint.  This need for a
repeated color label means that the combinatoric factor does not scale
as $N_c^4$ but only $N_c^3$, which, in combination with the $1/N_c$
factor from coupling constants, yields a contribution of
$N_c^2$---precisely as expected.

Reference~\cite{Cherman:2006iy} considered many classes of diagram.
In all cases the restrictions imposed by the condition that the color
combinations of the initial quarks and the final quarks must be
identical yielded overall contributions no higher than $O(N_c^2)$.
While these results greatly strengthen one's confidence that the mass
does indeed scale as $N_c^2$, Ref.~\cite{Cherman:2006iy} nevertheless
provided no general approach to prove that this result holds for all
classes of diagram; each class was analyzed separately.  The purpose
of this paper is to remedy this situation by providing a general
demonstration that the contribution of any $n$-quark cluster to the
baryon mass in QCD$_{\rm AS}$ scales as $N_c^2$ (or less) for any
class of diagram.

Our strategy is to introduce a diagrammatic simplification by
replacing diagrams of arbitrary complexity with effective diagrams
that preserve the connectivity between incoming and outgoing lines and
the $N_c$ counting of the original diagram.  These effective diagrams
resemble traffic circles or roundabouts.  Thus, despite the fact that
this approach is quite direct, it is aptly described as a ``roundabout
approach.''  In what follows we first illustrate the power of the
roundabout approach for the simple and known case of baryons in
QCD$_{\rm F}$ (Sec.~\ref{sec:QCD_F}), and then turn to the case of
interest, QCD$_{\rm AS}$ (Sec.~\ref{sec:QCD_AS}).
 
In the following demonstrations, we draw representative Feynman
diagrams.  When it is important to illustrate the color flow, we
follow 't~Hooft and use diagrams in which gluons appear as two
oppositely directed color lines.  Quarks in QCD$_{\rm F}$ are
represented by single color lines, while quarks in QCD$_{\rm AS}$ are
represented by doubled color lines pointing in the same direction, in
order to reflect the fact that quarks now carry two color indices.
The double-line representation for quarks in QCD$_{\rm AS}$ is used in
both Feynman diagrams and color-flow diagrams.  Whenever necessary,
colors are labeled with lower-case letters.

\section{Baryons in Fundamental QCD} \label{sec:QCD_F}

Each QCD$_{\rm F}$ quark carries one color index, and an operator in
QCD$_{\rm F}$ carrying unit baryon quantum number contains $N_c$
quarks, each with a distinct color.  Consider a general Feynman
diagram with $n$ quarks connected irreducibly ({\it i.e.}, one cannot
obtain two distinct interacting subdiagrams by cutting all of the
quark lines without also cutting a gluon line).  As demonstrated by
't~Hooft, diagrams with nonplanar gluons are suppressed, as are quark
loops.  Furthermore, neither connecting additional gluons to
pre-existing gluons nor connecting additional quarks increases the
$N_c$ counting.  All irreducible diagrams built this way are known to
scale as $N_c^1$~\cite{Witten:1979kh}.

\begin{figure}[ht]
    \centering
    \includegraphics[width=1.5in]{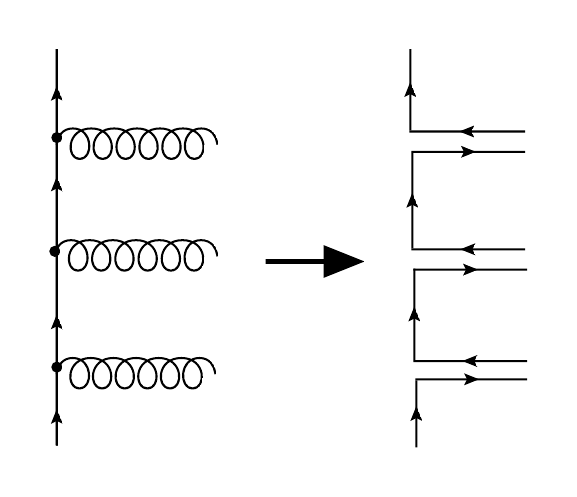}
    \caption{
    Multiple gluon connections to a single quark 
    and the corresponding color flow diagram.}
    \label{multipleconnections}
\end{figure}

\begin{figure}[ht]
    \centering
    \includegraphics[width=1.5in]{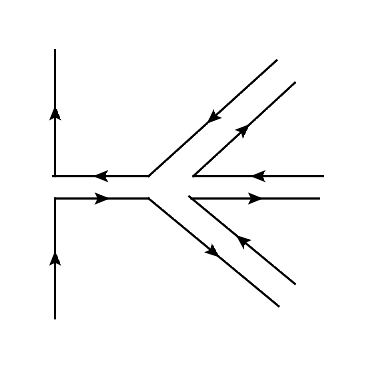}
    \caption{
    Equivalent color flow diagram for the case of multiple 
    gluon connections to a single quark given in
    Fig.~\ref{multipleconnections}.}
    \label{distortedlines}
\end{figure}

The only other diagrammatic possibility is the case of multiple gluon
connections to a single quark, an example of which is shown in
Fig.~\ref{multipleconnections}.  Notice that distorting the internal
quark lines gives a diagram (Fig.~\ref{distortedlines}) that, while
distinct as a Feynman diagram, is equivalent in terms of color flow to
one in which the quark effectively connects to only a single gluon,
with any additional gluons connected only to other gluons.

Since this color reduction can be done to any of the $n$ quarks in the
general Feynman diagram, the resulting equivalent diagram is one in
which a single gluon connects each of the $n$ quarks to a general
gluon diagram.  As demonstrated in Ref.~\cite{Witten:1979kh},
$n$-quark diagrams all scale as $N_c^1$ (or less).  If one now divides
by $N_c^n$ to remove the combinatoric factors associated with each
connected quark, one concludes that the gluon interaction must scale
as $1/N_c^{n-1}$.  The critical observation is that this $1/N_c^{n-1}$
scaling does not depend upon the details of the interaction.

An alternate method of deriving the $N_c$ dependence of the $n$-gluon
vertex follows from using the rescaled QCD Lagrangian,
\begin{eqnarray}
\lefteqn{{\cal L} = \frac{N_c}{g_s^2} } & & \nonumber \\
& & \times \left[ -\frac 1 4 \left( F_{\mu
\nu} \right)^a{}_{\! b} \left( F^{\mu \nu} \right)^b{}_{\! a} +
\overline{\psi}_a \left( i {D \! \! \! \! \slash} \right)^a{}_b \psi^b
- m \overline{\psi}_a \psi^a \right] \, , \nonumber \\
\end{eqnarray}
where the fields $A^\mu$ and $\psi$ each absorb a factor
$g_s/\sqrt{N_c}$ compared to the conventional definition.  This form
provides the most convenient starting point not only for proving the
finiteness of the large $N_c$ limit, but along the way shows that
3-point vertices scale as $1/\sqrt{N_c}$ and that the gluon 4-point
vertex scales as $1/N_c$.  Using the same approach, if the Lagrangian
is manipulated to allow for effective multi-gluon vertices (such as
appear in the passage from Fig.~\ref{multipleconnections} to
Fig.~\ref{distortedlines}), then the $n$-gluon vertex scales as
$N_c^{1-n/2}$.  Including the $n$ vertices where the gluons terminate
on quark lines brings in another $(1/\sqrt{N_c})^n$, making for a
complete interaction that scales as $N_c^{1-n}$, exactly as obtained
above.

\begin{figure}[ht]
    \centering
    \includegraphics[width=3.5in]{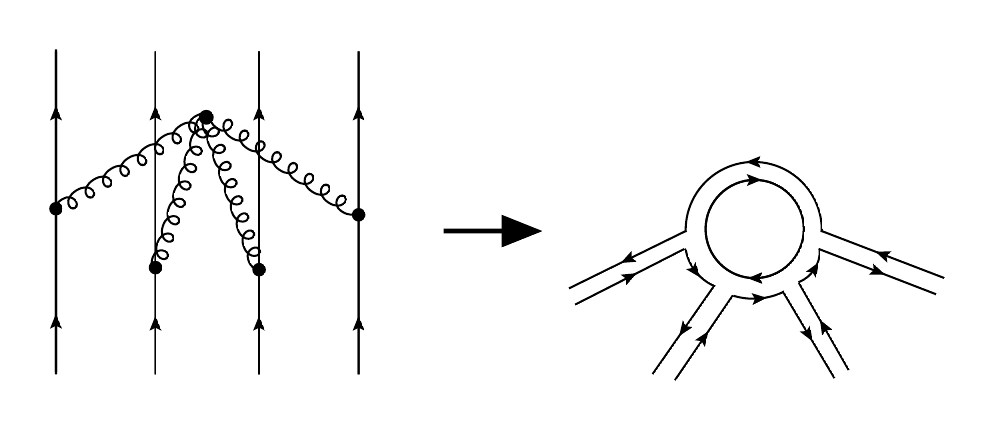}
    \caption{
    The $n$-gluon vertex and its corresponding representation as an 
    $n$-quark roundabout.}
    \label{roundabout}
\end{figure}

This observation motivates the introduction of the term {\it
roundabout}, or {\it traffic circle}, to depict such multi-gluon
interactions, in which each color line flows in at one location and
out at an adjacent location (which is allowed since the ordering of
the quark lines is unimportant).  As shown above, the $N_c$ counting
of this interaction only depends upon the number of quarks connected
to the diagram.  The most convenient depiction of the $n$-quark
roundabout is given in Fig.~\ref{roundabout}; its $N_c^{1-n}$
dependence is easily remembered by including a $1/N_c$ factor for each
incoming gluon, while the (artificial) color loop reminds one to
include the numerator factor of $N_c^1$.  It is worth noting that
't~Hooft's $N_c$ counting rules allow {\em any\/} arbitrarily
complicated $n$-point interaction among the gluons to be reduced to an
equivalent roundabout, so far as $N_c$ scaling is concerned.  In fact,
the roundabout represents a highly nonlocal QCD operator.

The initial step in reducing an $n$-quark diagram is therefore to make
all possible reductions of the form given in
Fig.~\ref{distortedlines}; color flow on any given quark line then
exits via a single equivalent gluon that ultimately enters a
roundabout.  With this notion in mind, it is easy to see that {\em
any\/} coupled quark cluster can be reduced to the form of
Fig.~\ref{roundabout} (in that specific case, for 4 quarks), with one
effective gluon leaving each quark and the effective gluons meeting at
a roundabout.  While such substitutions alter the Feynman diagrams and
hence the dynamics, they fully capture the color flow and hence the
$N_c$ counting.

With the idea of a roundabout in hand, generating Witten's argument
that $n$-quark Feynman diagrams scale as $N_c^1$ or less becomes
trivial.  Since $n$ quarks are connected, the combinatoric
contribution to the $N_c$ counting is $N_c^n$, while the interaction
contribution is at most 1/$N_c^{n-1}$.  Thus the total $N_c$ counting
is at most $N_c^n$/$N_c^{n-1}$, or simply $N_c$.

As seen in this section, the notion of the roundabout diagram
simplifies the derivation of the $N_c$ scaling for baryons in
QCD$_{\rm F}$.  However, the problem is sufficiently simple that the
construction is not really essential.  In the following section, we
apply the roundabout diagram approach to the more complicated case of
QCD$_{\rm AS}$, in which it plays a pivotal role to make the counting
tractable.

\section{Baryons in Antisymmetric QCD} \label{sec:QCD_AS}

Each QCD$_{\rm AS}$ quark has two color indices, and an operator in
QCD$_{\rm AS}$ creating a baryon contains $N_c(N_c \! - \! 1)/2 \sim
N_c^2$ quarks.  Consider a general Feynman diagram with $n$ connected
quarks.  We demonstrate that the interaction energy contribution to
the baryon mass scales at most as $N_c^2$ in the large $N_c$ limit.
It is therefore sufficient to show that all leading-order diagrams
scale as $N_c^2$.

\begin{figure}[ht]
    \centering
    \includegraphics[width=1in]{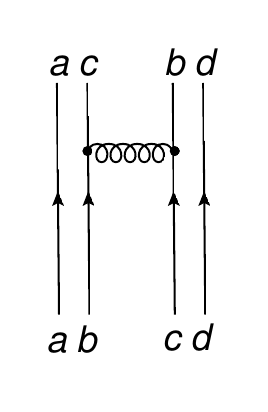}
    \caption{
    One-gluon exchange that does not contribute to color-singlet
    correlators in QCD$_{\rm AS}$ baryons.}
    \label{invalidexchange}
\end{figure}

\begin{figure}[ht]
    \centering
    \includegraphics[width=2in]{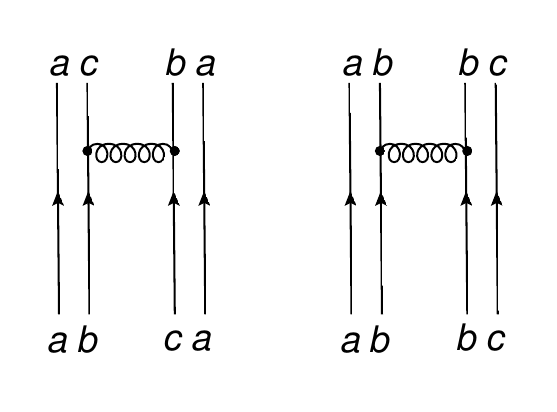}
    \caption{
    One-gluon exchanges that preserve QCD$_{\rm AS}$ quark color
    combinations and hence contribute to color-singlet correlators
    in QCD$_{\rm AS}$ baryons.}
    \label{validexchange}
\end{figure}

As noted in the Introduction, at first glance this scaling rule
appears to be violated by one-gluon exchange
(Fig.~\ref{invalidexchange}).  The four distinct color indices provide
a combinatoric contribution of $N_c^4$, while the interaction (with
two gluon vertices) scales as 1/$N_c$, thus giving a total dependence
of $N_c^3$.  However, as demonstrated in Ref.~\cite{Cherman:2006iy},
one-gluon exchange in QCD$_{\rm AS}$ does not contribute unless one
restricts the exchange to quarks that have one color index in common
(Fig.~\ref{validexchange}).  This constraint is necessary in order to
enforce Bolognesi's condition that a color-singlet baryon operator
contains a quark carrying any given pair of colors precisely
once~\cite{Bolognesi:2006ws}.

It is important to note that the one-gluon exchange in
Fig.~\ref{invalidexchange} remains physically valid even when all four
colors involved are distinct.  The Feynman diagram as given, with a
particular specified set of colors for the quarks, is perfectly
legitimate.  The key point is that, if the incident set of quarks as
drawn is part of a color singlet and the other quarks not pictured
maintain their color identities, then the final set does {\em not\/}
form a color singlet.  One obtains a color singlet by forming an
appropriate linear combination of quark color combinations, as appears
in the baryon operator.  However, color is conserved and moreover, the
color singlet contains each pair of colors once and only once.
One-gluon exchange diagrams that yield quarks with different color
combinations in the initial and final states can contribute only to
color-nonsinglet processes.  Summing over the initial color
combinations to yield a one-particle color-singlet state necessarily
causes such contributions to cancel completely.  The net contribution
of any Feynman diagram in which the set of color pairs of quarks in
the initial and final states differ must vanish {\em in correlation
functions of color-singlet operators}.

The upshot of this argument is that the set of color pairs of quarks
in the final state must be a permutation of the set in the initial
state in order to contribute to correlation functions of color-singlet
operators.  The only ways for one-gluon exchange to respect this
condition are either if the quarks share one color in common while
exchanging the other, or if the gluon couples to two quark lines
carrying the same color, as in Fig.~\ref{validexchange}.  Note,
however, that the restriction of having a pair of colors in common
reduces the combinatoric factor from $N_c^4$ to $N_c^3$.  Combining
this $N_c^3$ with the $1/N_c$ from the two coupling constants yields a
total scaling of $N_c^2$, as desired.

The question we consider is whether the restriction that all
leading-order contributions to correlation functions of color-singlet
operators require the sets of quark color pairs in the initial and
final states to be the same up to permutation is sufficient to ensure
that all classes of diagram contribute no faster than $N_c^2$.  In
Ref.~\cite{Cherman:2006iy}, many classes of diagrams were considered,
and in fact all contributed at $O(N_c^2)$ or less.  The question we
address here is whether this result is completely general, holding for
any diagram and with any number of repeated color indices.  We claim
that it does and provide a general demonstration below.

Before addressing this general question, let us note that one's naive
intuition about which classes of diagrams can contribute at leading
order can easily be wrong.  Consider for example the role played by
interactions of quarks that have a color in common among the pair of
colors specifying each quark.  Naively it might seem that such
interactions are unimportant at large $N_c$.  After all, the odds that
any two quarks share a common color is $\sim \! 1/N_c$, and one might
expect the effect of such interactions to be suppressed by a factor of
$1/N_c$.  However, as seen in the case of one-gluon exchange, such
interactions occur at leading order, {\it i.e.}, $O(N_c^2)$.  In the
case of one-gluon exchange the reason is rather clear: Quarks with one
color in common {\em do\/} appear at relative $O(1/N_c)$ compared to
the typical case with all distinct colors; however, the typical case
happens not to contribute to color singlets.  In a similar way, one
might think that gluons that change the color of the quark line with
which they interact necessarily dominate over those that do not ({\it
i.e.}, those that are diagonal in color space), since they outnumber
the color-diagonal ones by a factor of $N_c$.  We refer to the latter
as {\it Cartan gluons\/} since they are associated with the generators
of the Cartan subgroup.  Again one finds that Cartan gluons can also
contribute at leading order, $N_c^2$.

Of course, the critical question is not whether one can find
leading-order contributions, but rather whether one can show that no
``superleading'' contributions scaling faster than $N_c^2$ occur.
However, the fact that repeated color indices and Cartan gluons
contribute at leading order means that they are as important as the
apparently more typical examples.  Hence, they need to be included in
a demonstration that no superleading contributions exist. This fact
greatly complicates the analysis.

The strategy to demonstrate that no class of diagram scales faster
than $N_c^2$ has three main steps.  The first two remove certain types
of gluons from the diagram, which breaks the set of interacting quarks
into smaller clusters of interacting quarks.  We show that the leading
$N_c$ scaling of the contribution to a color-singlet correlator of the
combined diagram is no larger than the $N_c$ scaling of the
contribution of the smaller clusters.  We call the smallest of these
{\it irreducible clusters}.  The final step exploits the roundabout
diagrams to show that irreducible clusters contribute at $O(N_c^2)$ or
less.

\begin{figure}[ht]
    \centering
    \includegraphics[width=2.9in]{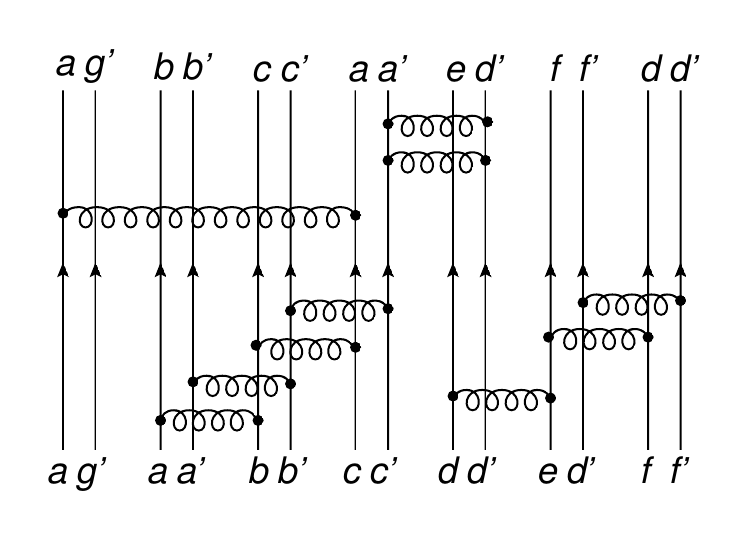}
    \caption{
    A leading-order generic Feynman diagram in QCD$_{\rm AS}$
    contributing to color-singlet correlators. } 
    \label{generaldiagram}
\end{figure}

We illustrate the procedure on a particular diagram
(Fig.~\ref{generaldiagram}); however, all of the individual steps
apply to any arbitrarily complicated diagram.  One can check that this
diagram does indeed contribute, since it preserves the color
combinations between initial and final states.  Furthermore, by
counting the number of distinct indices (12) and the number of gluon
vertices (20), this diagram evidently scales as $N_c^{12}$/$N_c^{10}$
= $N_c^2$, making it a leading-order diagram.  While obtaining this
result is simple in this particular case, the key question is how to
break up the generically complicated combinatoric problem into pieces
that can be simply analyzed.

\begin{figure}[ht]
    \centering
    \includegraphics[width=3in]{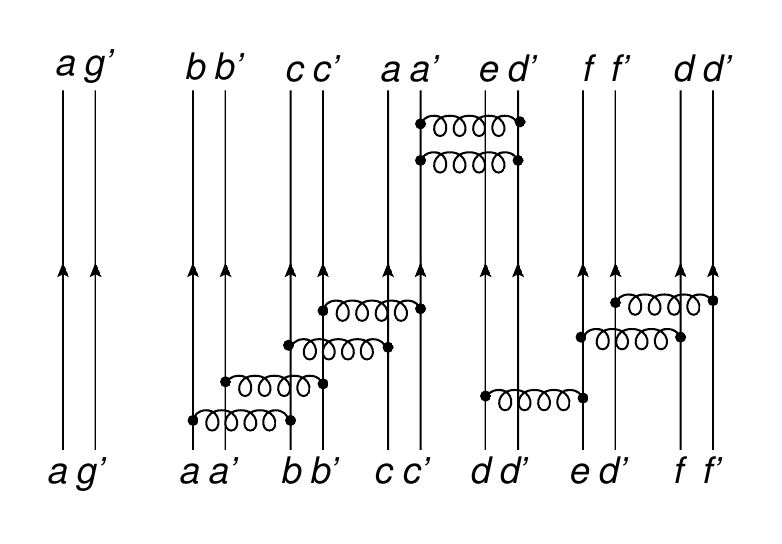}
    \caption{
    The Feynman diagram of Fig.~\ref{generaldiagram} after the first
    step, the removal of Cartan gluons.  Two mutually noninteracting
    clusters remain.}
    \label{clusters}
\end{figure}

The first step of the simplification is to disconnect the general
diagram by removing any Cartan gluons, keeping track of the number
removed and where they were connected.  Since this type of connection
results in a color line that interacts but does not change its color
value, the removal of Cartan gluons can completely separate clusters
of quarks without changing color flow (In the case of QCD$_{\rm AS}$,
one of the two color lines of a particular quark is allowed not to
interact as long as the other one does, thus coupling the whole quark
to the diagram).  In our example, this step decouples only the $a g'$
quark from the diagram (Fig.~\ref{clusters}).  Notice that if an
interaction occurred between the $d'$ from the $d d'$ quark and the
$d'$ from the $e d'$ quark, this Cartan gluon would also have been
removed in this step.  However, such an extra interaction was not
included because it would have clearly made the diagram subleading in
$1/N_c$.  The distinction is that this gluon does not connect to a
quark in the diagram that carries a noninteracting line (such as $g'$)
that provides a combinatoric factor $N_c$ to compensate the additional
vertex $1/\sqrt{N_c}$ factors.

This step is important because all of the remaining gluons change
quark colors, allowing one to develop a connection between color flow
and $N_c$ scaling.  The removal of a Cartan gluon either separates the
diagram into two distinct interacting clusters or it does not.  If
not, then the diagram was necessarily subleading: An identical diagram
without it would have the same color flow and combinatorics but one
pair fewer of coupling constants and hence one factor less of $1/N_c$.
One can thus dismiss this case as uninteresting in establishing
whether or not any class of diagrams contributes at a superleading
order.  If the removal breaks the diagram into two clusters, then
suppose further that one can show the contribution from any connected
cluster not containing a Cartan gluon scales no faster than $N_c^2$.
In that case, one easily sees that connecting two such clusters also
leads to $N_c^2$ scaling: Each cluster contributes $N_c^2$, the two
coupling constants together contribute $1/N_c$, and a combinatoric
suppression factor of $1/N_c$ combines with them to give a total
scaling of $N_c^2$.  The origin of this last factor is easy to see: In
order for the two clusters to be coupled by a Cartan gluon, they must
share a color, suppressing the combinatoric possibilities compared to
two clusters with completely distinct colors.  Of course, this
argument is completely general and does not require the diagram to
have the form of Fig.~\ref{generaldiagram}.  The problem is thus
reduced to showing that graphs containing no Cartan gluons scale as
$N_c^2$ or less.

\begin{figure}[ht]
    \centering
    \includegraphics[width=3in]{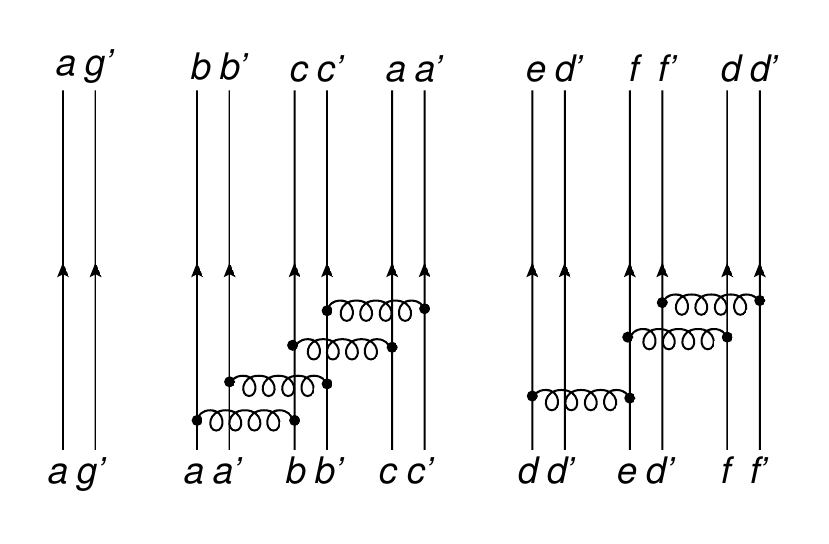}
    \caption{
    The Feynman diagram of Fig.~\ref{generaldiagram} after the second
    step, the separation of regular clusters into irreducible
    clusters.  Three irreducible clusters remain.}
    \label{irreducibleclusters}
\end{figure}

The clusters created in the first step above are now analyzed
individually.  Each cluster may further consist of some number of
separate quark groups, each of which shares no color indices with
another group, nevertheless connected to one another (by means of
colors passed back and forth between the two).  The second step is to
remove the gluons that connect these groups, again keeping track of
the number removed and where they were connected.  This action results
in a set of what we call {\it irreducible clusters}, irreducible
because no quark can be removed since each quark interacts, changes
color, and is coupled to every other quark in a sequence.  In our
example, the cluster consisting only of the $a g'$ quark is itself an
irreducible cluster, while the other cluster can be separated into two
irreducible clusters by removing the two gluons connecting the $c c'$
quark to the $d d'$ quark (Fig.~\ref{irreducibleclusters}).

It is easy to see that the leading scaling of a cluster made up of
irreducible clusters must be $N_c^2$, provided that an irreducible
cluster also scales as $N_c^2$.  By construction, the gluons
connecting two irreducible clusters must first shift color from one
irreducible cluster to the other and then back, which requires at
least two gluon exchanges, or four coupling constants, which
contribute a factor of $1/N_c^2$.  Combining with a factor of $N_c^2$
from each irreducible cluster yields $N_c^2$, as required.  A simple
inductive argument shows that one can then combine any number of
irreducible clusters together in just this manner while retaining the
$N_c^2$ scaling behavior.  Thus the problem is now reduced to showing
that irreducible clusters scale as $N_c^2$.

The third and final step is the demonstration that any irreducible
cluster, regardless of its size and the details of its interactions,
scales at most as $N_c^2$.  Roundabout diagrams greatly aid in this
step.

Note that, in all of the irreducible clusters of
Fig.~\ref{irreducibleclusters}, the two color lines in each quark are
divided into two classes: those with an unprimed letter written on the
left (L indices) and those with a primed letter written on the right
(R indices).  This distinction is {\it a priori\/} artificial, since
the quarks are antisymmetric in the indices.  It is noteworthy,
however, that the gluons in Fig.~\ref{irreducibleclusters} only
connect among quark L indices or among quark R indices, meaning that L
indices of the various quarks exchange colors exclusively amongst
themselves, and similarly for R indices.  Thus, while the question of
which index one labels as L or R is arbitrary, the connections break
up into two distinct classes of connected color flow for the type of
diagram seen here.  Of course, one can consider diagrams where the
color flow does {\em not\/} break up into two separate classes.
However, it can be shown that such graphs are necessarily subleading.
In the following argument we assume that this is true, deferring until
the end a demonstration, for even the case of irreducible diagrams
with separate color flow among the L and R indices remains more
complicated than the situation in QCD$_{\rm F}$.

A principal reason for this complication is that an irreducible
cluster can still have any number of repeated L or R indices.
However, it is possible to simplify the analysis by exploiting the
$n$-quark roundabout diagrams corresponding to each interaction.  In
order to describe quark lines that do not interact, we consider a
single noninteracting quark line to form a trivial roundabout (or cul
de sac; see Fig.~\ref{roundabouts} for examples).  Although more than
one roundabout can appear in an arbitrary diagram, the roundabouts
themselves look exactly like those in the case of QCD$_{\rm F}$.  The
L indices and the R indices interchange separately, and for each of L
and R one requires one or more roundabouts (some of which may be
trivial) to describe any interaction.

We now show that the leading-order diagrams maximize the number of
complete roundabouts for a given number of repeated indices.  This
crucial feature follows from the fact that replacing one roundabout
with two (whenever it is valid to do so, given the color connections
between quarks required by the gluons present in the original diagram)
effectively contributes an additional factor of $N_c$.  When
roundabouts split, an additional color line can interact (introducing
its combinatoric factor of $N_c$) without the cost of an additional
factor of $1/N_c$.  In other words, a roundabout with $n \! = \! n_1
\! + \! n_2$ lines scales as $N_c^{1 - n_1 - n_2}$, while the
combination of roundabouts with $n_1$ and $n_2$ lines scales as
$N_c^{1 - n_1} N_c^{1 - n_2} \! = \! N_c^{2 - n_1 - n_2}$.  A
roundabout with repeated indices is subleading compared to one in
which the repeated indices are separated into two roundabouts, because
the former loses a combinatoric factor of $N_c$ compared to the
latter.  In summary, a leading-order diagram is one in which none of
the roundabouts contain more than one instance of a given index,
whether or not that index is repeated elsewhere.  In our example, we
have drawn the roundabout diagrams corresponding to each irreducible
cluster (Fig.~\ref{roundabouts}).  The first irreducible cluster
consists of two trivial roundabouts, one from L lines and one from R
lines.  The second consists of two ordinary roundabouts, one L and one
R.  The third consists of one ordinary roundabout from the L lines and
one ordinary plus one trivial roundabout from the R lines.

\begin{widetext}
\begin{center}
\begin{figure}[ht]
    \includegraphics[width=6.0in]{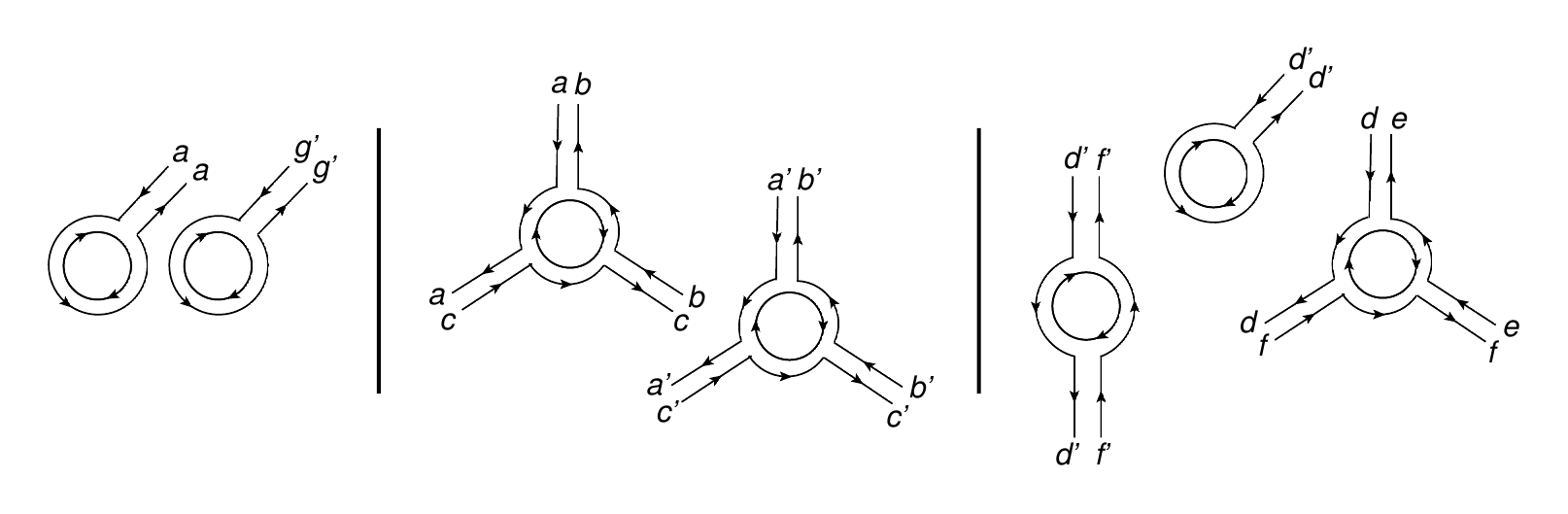}
    \caption{
    The L and R roundabout diagrams for each of the 
    three irreducible clusters in Fig.~\ref{irreducibleclusters}.}
    \label{roundabouts}
\end{figure}
\end{center}
\end{widetext}

We now demonstrate that the number of possible complete roundabouts
has an upper limit $N_r \! + \! 2$, where $N_r$ is the number of
redundant indices (separately counted for L and for R lines), by which
we mean the total number of color indices minus the number of distinct
color values carried by the indices.  We then show how this result
implies that no irreducible cluster can scale with $N_c$ faster than
$N_c^2$.

The simplest case (in which each quark line has a different color
index, and all of the indices permute) is composed of two roundabouts,
one encompassing all L indices and one encompassing all R indices; it
is consistent with the general result because $N_r \! = \! 0$.  Now
consider a general irreducible cluster with some number $N_r$ of
redundant indices.  By turning a previously distinct color index into
a repeat of one already present, one creates a diagram with $N_r \! +
\! 1$ redundant indices.  As argued above, the diagram gains a power
of $N_c$ if the roundabout in which this new redundancy occurs is
replaced with two smaller ones, for a gain of one roundabout;
repeating this argument for each of the $N_r$ redundant indices shows
that leading-order diagrams occur when the number of roundabout
interactions is $N_r \! + \! 2$.

With this result in hand, we now directly compute the maximum $N_c$
dependence of any irreducible cluster.  Since two color indices are
provided by each AS quark, and each distinct color index adds a
combinatoric factor of $N_c$, the combinatoric contribution is
$N_c^{2n-N_r}$, where $n$ is the number of quarks connected in the
irreducible cluster.  Each roundabout scales as $N_c^{1-n_i}$, $i
\! = \! 1, 2, \ldots, N_r \! + \! 2$, where $n_i$ is the number of
color lines appearing in the $i^{\rm th}$ roundabout, and each of the
$n$ quarks (which has an L and an R color line) contributes to two
roundabouts.  The product of these factors over the $N_r \! + \! 2$
roundabouts is just $N_c^{N_r + 2 - 2n}$.  Thus the total $N_c$
dependence is $N_c^{2n - N_r} \! \times N_c^{N_r + 2 - 2n}$, which
reduces to just $N_c^2$.  Of course{\bf ,} the $N_c$ scaling can be
less than this value if the total number of roundabouts does not reach
the upper limit of $N_r \! + \! 2$.

In our example (Fig.~\ref{roundabouts}), the first irreducible
cluster consists of two trivial roundabouts, each with an
interaction contribution $1/N_c^0$ = $N_c^0$ and a combinatoric
contribution $N_c^1$.  Thus the total $N_c$ scaling is indeed $N_c^2$.
The second irreducible cluster also scales as $N_c^2$, with
combinatoric contribution $N_c^6$ and interaction contribution
$1/N_c^4$.  The third irreducible cluster has combinatoric
contribution $N_c^5$ and interaction contribution $1/N_c^3$, for a
total $N_c$ dependence of $N_c^2$.

The basic demonstration that no class of diagram scales faster than
$N_c^2$ is now complete.  However, it was based on the assumption that
no leading-order diagrams mix L and R indices.  To prove this claim,
consider any connected diagram among $n$ quarks featuring an
interaction between the two sets---namely, an LR gluon---and then
consider the connected $n$-quark diagram in which the LR gluons are
simply deleted.  Note that the diagram remains connected because an LR
mismatch does not appear unless both the L line of the second quark
and the R line of the first quark are required to connect to other
quark lines of the corresponding handednesses elsewhere in the
diagram.  However, deleting a given LR gluon line gains a power of
$N_c$ from the two now-absent trilinear vertices, and since every
quark remains connected to the diagram, this deletion loses no
combinatoric factors of $N_c$.  Thus, the effect of deleting the LR
gluons is to increase the $N_c$ scaling, indicating that the original
graph was subleading.

One might worry that possible factors of $N_c$ could also be lost in
this deletion, arising from color loops that the LR gluon might have
completed.  However, the act of reducing all gluon interactions to
roundabouts means that all color lines enter the diagram at an initial
point, pass through a roundabout, and leave through a final point.
Reminding the reader that the color loop in each roundabout diagram is
merely a representational device, one sees that no color loops remain
in a diagram converted to roundabout form, and hence no loop factors
of $N_c$ are lost in this deletion.  In fact, one can further show,
once the Cartan gluons are removed, that diagrams with LR gluons are
suppressed by a factor of at least $N_c^{-2}$ compared to diagrams
without them: Since non-Cartan gluons exchange colors on the quark
lines, exchanging color values between the L and R sides and switching
them back again requires at least two LR gluon exchanges, and hence
induces a suppression of $1/N_c^2$ compared to diagrams without this
color exchange.  One concludes that the set of diagrams with L color
lines and R color lines separated always includes diagrams with the
leading-order $N_c$ counting.  As mentioned above, this separation is
present in our example.

Thus we have completed our demonstration that no class of connected
$n$-quark diagrams contributes to color-singlet correlators with power
greater than $N_c^2$ (provided the number of gluons in the graph is
finite).  Following the arguments of
Refs.~\cite{Witten:1979kh,Luty:1993fu}, this conclusion implies that
the QCD$_{\rm AS}$ baryon mass scales as $N_c^2$.  While the arguments
sketched here are informal, they could be converted into a rigorous
theorem with no essential difficulty.  However, one should recall that
the arguments in Refs.~\cite{Witten:1979kh,Luty:1993fu} are themselves
not rigorous: They depend upon the assumption that one can deduce the
baryon mass scaling behavior from the behavior of Feynman diagrams,
which are intrinsically perturbative in nature.  While extremely
unlikely, it is logically possible that fundamentally nonperturbative
effects not diagrammatically expressible could alter this result.  The
conclusion one reaches from such considerations is that the claim that
the baryon mass scales as $N_c^2$ in QCD$_{\rm AS}$ is now as
solid as the claim that the baryon mass scales as $N_c^1$ in QCD$_{\rm
F}$.

\vskip 3ex
{\it Acknowledgements.}  We thank the organizers and participants of
the INT workshop ``New Frontiers in Large $N$ Gauge Theories,'' where
this work was initiated, and also the INT and the University of
Washington for their hospitality.  T.D.C.\ acknowledges the support of
the U.S.\ Dept.\ of Energy under Grant No.\ DE-FG02-93ER-40762.
R.F.L.\ acknowledges the support of the NSF under Grant No.\
PHY-0757394.


\begin{thebibliography}{99}

%\cite{'tHooft:1973jz}
\bibitem{'tHooft:1973jz}
  G.~'t~Hooft,
  %``A PLANAR DIAGRAM THEORY FOR STRONG INTERACTIONS,''
  Nucl.\ Phys.\  B {\bf 72}, 461 (1974).
  %%CITATION = NUPHA,B72,461;%%

%\cite{'tHooft:1974hx}
\bibitem{'tHooft:1974hx}
  G.~'t~Hooft,
  %``A Two-Dimensional Model For Mesons,''
  Nucl.\ Phys.\  B {\bf 75}, 461 (1974).
  %%CITATION = NUPHA,B75,461;%%

%\cite{Witten:1979kh}
\bibitem{Witten:1979kh}
  E.~Witten,
  %``Baryons In The 1/N Expansion,''
  Nucl.\ Phys.\  B {\bf 160}, 57 (1979).
  %%CITATION = NUPHA,B160,57;%%

\bibitem{emergent}
%\cite{Gervais:1983wq}
%\bibitem{Gervais:1983wq}
  J.L.~Gervais and B.~Sakita,
  %``Large N QCD Baryon Dynamics: Exact Results From Its Relation To
  %The Static Strong Coupling Theory,''
  Phys.\ Rev.\ Lett.\  {\bf 52}, 87 (1984);
  %%CITATION = PRLTA,52,87;%%
%\cite{Gervais:1984rc}
%\bibitem{Gervais:1984rc}
%  J.L.~Gervais and B.~Sakita,
  %``Large-N baryonic soliton and quarks,''
  Phys.\ Rev.\  D {\bf 30}, 1795 (1984).

\bibitem{DM}
%\cite{Dashen:1993as}
%\bibitem{Dashen:1993as}
  R.F.~Dashen and A.V.~Manohar,
  %``Baryon - pion couplings from large N(c) QCD,''
  Phys.\ Lett.\  B {\bf 315}, 425 (1993)
  [arXiv:hep-ph/9307241];
  %%CITATION = PHLTA,B315,425;%%
%\cite{Dashen:1993ac}
%\bibitem{Dashen:1993ac}
%  R.F.~Dashen and A.V.~Manohar,
  %``1/N(c) corrections to the baryon axial currents in QCD,''
%  Phys.\ Lett.\
  B {\bf 315}, 438 (1993)
  [arXiv:hep-ph/9307242].
  %%CITATION = PHLTA,B315,438;%%

%\cite{Dashen:1993jt}
\bibitem{Dashen:1993jt}
  R.F.~Dashen, E.E.~Jenkins and A.V.~Manohar,
  %``The 1/N(c) expansion for baryons,''
  Phys.\ Rev.\  D {\bf 49}, 4713 (1994)
  [Erratum-ibid.\  D {\bf 51}, 2489 (1995)]
  [arXiv:hep-ph/9310379].
  %%CITATION = PHRVA,D49,4713;%%

%\cite{Carone:1993dz}
\bibitem{Carone:1993dz}
  C.~Carone, H.~Georgi and S.~Osofsky,
  %``On spin independence in large N(c) baryons,''
  Phys.\ Lett.\  B {\bf 322}, 227 (1994)
  [arXiv:hep-ph/9310365].
  %%CITATION = PHLTA,B322,227;%%

%\cite{Dashen:1994qi}
\bibitem{Dashen:1994qi}
  R.F.~Dashen, E.E.~Jenkins and A.V.~Manohar,
  %``Spin-Flavor Structure of Large N Baryons,''
  Phys.\ Rev.\ D {\bf 51}, 3697 (1995)
  [arXiv:hep-ph/9411234].
  %%CITATION = PHRVA,D51,3697;%%

%\cite{Jenkins:1994md}
\bibitem{Jenkins:1994md}
  E.E.~Jenkins and A.V.~Manohar,
  %``Baryon Magnetic Moments in the 1/N Expansion,''
  Phys.\ Lett.\  B {\bf 335}, 452 (1994)
  [arXiv:hep-ph/9405431].
  %%CITATION = PHLTA,B335,452;%%

%\cite{Luty:1994ub}
\bibitem{Luty:1994ub}
  M.A.~Luty, J.~March-Russell and M.J.~White,
  %``Baryon magnetic moments in a simultaneous expansion in 1/N and
  %m(s),''
  Phys.\ Rev.\  D {\bf 51}, 2332 (1995)
  [arXiv:hep-ph/9405272].
  %%CITATION = PHRVA,D51,2332;%%

%\cite{Jenkins:1995td}
\bibitem{Jenkins:1995td}
  E.E.~Jenkins and R.F.~Lebed,
  %``Baryon Mass Splittings in the 1/N_c Expansion,''
  Phys.\ Rev.\ D {\bf 52}, 282 (1995)
  [arXiv:hep-ph/9502227].
  %%CITATION = PHRVA,D52,282;%%

%\cite{Dai:1995zg}
\bibitem{Dai:1995zg}
  J.~Dai, R.F.~Dashen, E.E.~Jenkins and A.V.~Manohar,
  %``Flavor Symmetry Breaking in the 1/N Expansion,''
  Phys.\ Rev.\  D {\bf 53}, 273 (1996)
  [arXiv:hep-ph/9506273].
  %%CITATION = PHRVA,D53,273;%%

%\cite{Corrigan:1979xf}
\bibitem{Corrigan:1979xf}
  E.~Corrigan and P.~Ramond,
  %``A Note On The Quark Content Of Large Color Groups,''
  Phys.\ Lett.\  B {\bf 87}, 73 (1979).
  %%CITATION = PHLTA,B87,73;%%

%\cite{Armoni:2003gp}
\bibitem{Armoni:2003gp}
  A.~Armoni, M.~Shifman and G.~Ve\-ne\-zi\-a\-no,
  %``Exact results in non-supersymmetric large N orientifold field
  % theories,''
  Nucl.\ Phys.\  B {\bf 667}, 170 (2003)
  [arXiv:hep-th/0302163];
  %%CITATION = NUPHA,B667,170;%%
%\cite{Armoni:2003fb}
%\bibitem{Armoni:2003fb}
  % A.~Armoni, M.~Shifman and G.~Veneziano,
  %``SUSY relics in one-flavor QCD from a new 1/N expansion,''
  Phys.\ Rev.\ Lett.\  {\bf 91}, 191601 (2003)
  [arXiv:hep-th/0307097];
  %%CITATION = PRLTA,91,191601;%%
%\cite{Armoni:2004uu}
%\bibitem{Armoni:2004uu}
%  A.~Armoni, M.~Shifman and G.~Veneziano,
  %``From super-Yang-Mills theory to QCD: Planar equivalence and its
  %implications,''
  arXiv:hep-th/0403071.
%%CITATION = HEP-TH/0403071;%%

%\cite{Cherman:2009fh}
\bibitem{Cherman:2009fh}
  A.~Cherman, T.D.~Cohen and R.F.~Lebed,
  %``All you need is N: Baryon spectroscopy in two large N limits,''
  Phys.\ Rev.\  D {\bf 80}, 036002 (2009)
  [arXiv:0906.2400 [hep-ph]].
  %%CITATION = PHRVA,D80,036002;%%

%\cite{Luty:1993fu}
\bibitem{Luty:1993fu}
  M.A.~Luty and J.~March-Russell,
  %``Baryons from quarks in the 1/N expansion,''
  Nucl.\ Phys.\  B {\bf 426}, 71 (1994)
  [arXiv:hep-ph/9310369].
  %%CITATION = NUPHA,B426,71;%%

%\cite{Bolognesi:2006ws}
\bibitem{Bolognesi:2006ws}
  S.~Bolognesi,
  %``Baryons and Skyrmions in QCD with Quarks in Higher
  % Representations,''
  Phys.\ Rev.\ D {\bf 75}, 065030 (2007)
  [arXiv:hep-th/0605065].
  %%CITATION = PHRVA,D75,065030;%%

%\cite{Cherman:2006iy}
\bibitem{Cherman:2006iy}
  A.~Cherman and T.D.~Cohen,
  %``The skyrmion strikes back: Baryons and a new large N(c) limit,''
  JHEP {\bf 0612}, 035 (2006)
  [arXiv:hep-th/0607028].
  %%CITATION = JHEPA,0612,035;%%

\end{thebibliography}
\end{document}